\documentclass[11pt,twoside]{article}

\usepackage{asp2006}
\usepackage{epsf}
\usepackage{lscape}
\usepackage{natbib}

\begin{document}

\title{The Origin of Wavelength-Dependent Continuum Delays in AGNs -- a New Model}

\author{C. Martin Gaskell}
\affil{Department of Physics and Astronomy, University of Nebraska,
\\Lincoln, NE 68588, USA}

\begin{abstract}
A model of wavelength-dependent lags in optical continuum
variability of AGNs is proposed which avoids the problems of the
popular ``lamp-post'' model. Rather than being due to reprocessing
of high-energy radiation from a hypothetical source above the
accretion disk, the wavelength-dependent delays observed from the B
to I bands are instead due to contamination of an intrinsically
coherently variable continuum with the Wien tail of the thermal
emission from the hot dust in the surrounding torus.  The new model
correctly gives the size, wavelength dependence, and luminosity
dependence of the lags, and quantitatively predicts observed color
hysteresis. The model also explains how the measured delays vary
with epoch of observation. There must also be contamination by
scattered light and this can be detected by a lag in the polarized
flux.
\end{abstract}

\section{Introduction}

The timescale of wavelength-dependent lags in AGNs has long been
problematic since the delays are much longer than a dynamical
timescale. When delays were first convincingly established in
NGC~7469 \citep{wanders97,collier98} we interpreted them as a
consequence of light-travel time in the external illumination of a
disk with a $T \propto R^{-3/4}$ radial temperature structure.  This
gives delays, $\tau \propto \lambda^{4/3}$ \citep{collier98,
kriss00}. From broad-band optical photometry, \citet{sergeev05} have
found many more wavelength-dependent lags and made the important
discovery that the lags are luminosity-dependent with $\tau \propto
L^{1/2}$. To explain this they postulate that the height of the
external illumination source depends on the square-root of the
luminosity.

\section{Problems with Lamp Posts}

Although the external-illumination (``lamp-post'') model can readily
reproduce the wavelength dependence of the lags in NGC~7469, I
believe it has enormous problems. The first major problem is that
the ``lamp'' is not seen at {\it any} wavelength. It shines on the
disk but never in the direction of the observer.  In the terminology
of the {\it International Dark Sky Association} the lamp is a
``fully-shielded'' fixture! While observational optical astronomers
consider this to be highly desirable for all fixtures on terrestrial
lamp posts, this is impossible for the putative external sources of
illumination above AGN accretion disks -- no shield could survive in
the harsh conditions near the hypothetical energy source. A second
major problem is that after the correct subtraction of the host
galaxy light, the UV and optical variability amplitude is large --
an order of magnitude is not unusual.  To explain the lags with
external illumination requires that this order-of-magnitude
variability be due to the external illumination. If this is so, the
luminosity of the ``lamp'' exceeds that of the accretion disk, the
disk is irrelevant, and the lamp-post model is inconsistent, since
it requires the disk radial temperature dependence to be determined
by the accretion disk.

\begin{figure}[t!]
  \vskip 0.2cm
  \centering
  \epsfxsize=10cm
  \epsfbox{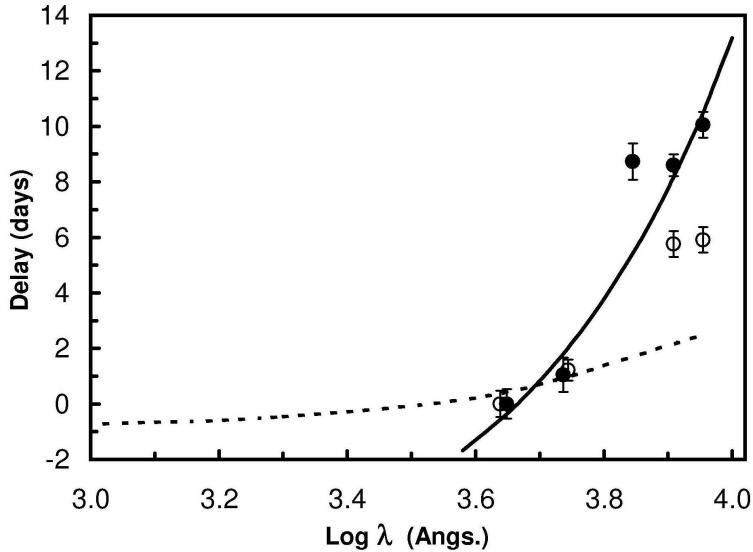}
  \caption{Normalized wavelength dependent delays scaled to the size of NGC~5548.  Filled circles are from the centroids of the
  cross-correlation functions; open circles are from the peaks of the cross-correlation functions.  The solid line
  is a fit of $\tau \propto \lambda^{4/3}$ through the mean centroid delays observed at $\lambda\lambda$ 4400, 5500,
  8000, and 9000.  The dotted line is a $\tau \propto \lambda^{4/3}$ relationship
  fit to the observed delay between $\lambda$1350 and $\lambda$5100 in NGC~5548.}
  \label{fig1}
\end{figure}

In Fig.~\ref{fig1} I show the mean lags relative to the B band
($\lambda$4400) for the 14 AGNs measured by \citet{sergeev05}.  The
lags were determined with the cross-correlation function (CCF)
technique of \cite{gaskell86}. Since the lags are luminosity
dependent, I have normalized the lags from the centroids of the CCFs
of the $\lambda$8000 and $\lambda$9000 bands to the corresponding
\citet{sergeev05} centroid lags for NGC~5548. The error bars give
errors in the means.  It can be seen that there is a clear
wavelength dependence and that this can easily be fit by $\tau
\propto \lambda^{4/3}$. However, this fit (the solid line in
Fig.~\ref{fig1}) predicts a far larger UV-to-optical lag than is
observed.  The dotted line shows a $\tau \propto \lambda^{4/3}$ fit
to the actual $\lambda$1350--$\lambda$5100 delay we reported for
NGC~5548 \citep{korista95}.

\section{The Effect of Optical Emission from the Dusty Torus}

The sharp increase in lag at long optical wavelengths in
Fig.~\ref{fig1} has no explanation in the lamp-post model. I propose
instead that it is due to contamination from optical emission from
the hot dust in the inner torus. Although the dust emission peaks in
the IR, the high dust sublimation temperature ($\sim 1500$\,K) means
that there is substantial {\it optical} emission as well\footnote{A
candle flame is at the graphite/PAH condensation temperature and a
candle emits in the optical!}.

\begin{figure}[t!]
  \vskip 0.2cm
  \centering
  \epsfxsize=10cm
  \epsfbox{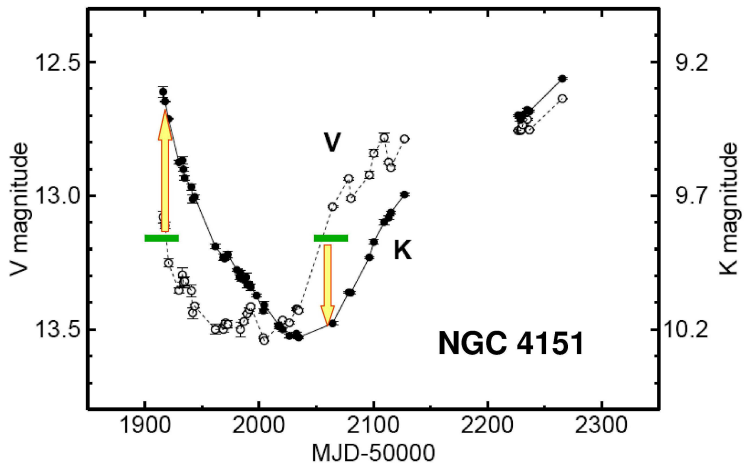}
  \caption{The K-band flux lagging the V-band flux in NGC~4151.  The vertical arrows show
  how the (V-K) colors are substantially different at different epochs because of the
  lag of the K-band flux. Adapted from \citet{minezaki06}}
  \label{fig2}
\end{figure}

It is well known that the variability of the IR dust emission lags
the optical variability (see Fig.~\ref{fig2}), and IR lags have been
determined for many objects \citep[see also these
proceedings]{suganuma06}.  The contaminating dust emission flux at
shorter wavelengths will thus also lag the direct emission from the
AGN. If two time series are cross correlated, the effect of
contaminating one series with a third series with a different lag is
to shift the peak in the CCF. This was modeled in a different
context by \citet{gaskell87}. In the model proposed here, the flux
in the R band, say, is the sum of the intrinsically-varying
continuum (assumed to be varying coherently) and a small, much
delayed contribution from the Wien tail of the hot dust.

The size of the lag primarily depends linearly on two things: (a)
the ratio of the contaminating flux to the intrinsic flux, and (b)
the inner radius of the dusty torus. These dependencies give us two
predictions: first, the optical lag will increase with increasing
wavelength (because the flux from the dust increases with
wavelength), as is shown to be the case in Fig.~\ref{fig1}. The
second prediction is that because the inner radius of the torus
(which is determined by the dust sublimation radius) increases as
$L^{1/2}$ \citep{suganuma06}, the relative lags will also increase
as $L^{1/2}$, as has been observed already by \citet{sergeev05}.

In Fig.~\ref{fig1} the R band ($\lambda$7000) lag lies significantly
above the line fit to the other points.  This is to be expected
because the strong H$\alpha$ emission line falls within the R
passband and so introduces additional delayed contamination.
\citet{korista01} have also pointed out that broad lines will
produce diffuse continuum emission.  This will be another source of
lagged contamination.

Because the dust emission comes from an extended region, its
variabilty is smeared out as well as delayed.  Rapid variations are
washed out. This gives two additional predictions. The first is that
the CCF will be asymmetric and the centroid of the CCF, which is
less sensitive to the variability power spectrum (see Koratkar \&
Gaskell 1991), will show a larger delay than the peak of the CCF.
It can be seen in Fig.~\ref{fig1} that this indeed the case. The
second prediction is that the lag given by the peak of the CCF will
be smaller when the variability is more rapid.  Thus different lags
will be measured at different times.

A final prediction is that there will be {\it hysteresis} in the
color-magnitude diagram. It is obvious in Fig.~\ref{fig2} that the
(V-K) colors are substantially different at two different epochs
with similar V flux levels.  In the model I have proposed, optical
to IR colors can be predicted from the V-band light curve alone. For
example, the hysteresis found by \citet{bachev04} in their (V-I)
versus V diagram for Mrk 279 is quantitatively reproduced.

The contamination model proposed here can be applied to other
wavelength regions (e.g., cross-correlation analyses of X-ray
variability). Because of the effects of contamination on
cross-correlation analyses it is important to note that the lag
given by a CCF often does {\it not} correspond to a physical scale.

In AGNs there must also be substantial contamination from {\it
scattered} radiation. This could be an explanation of the general
smoothness of optical light curves. Since the albedo of scatterers
is largely wavelength independent, this will not cause
wavelength-dependent lags, but the contamination can be detected
through lags in the polarized flux \citep{shoji05}.

\acknowledgments

This research has been supported by the National Science Foundation
through grant AST 03-07912 and by the Space Telescope Science
Institute through grant AR-09926.01.

\end{document}